\documentclass[twocolumn,pre,superscriptaddress]{revtex4}

\usepackage{dcolumn}
\usepackage{amsmath}

\usepackage{graphicx}
\usepackage{subfigure}

\renewcommand{\d}{\mathrm{d}}
\newcommand{\e}{\mathrm{e}}

\newcommand{\av}[1]{\langle#1\rangle}

\newcommand{\defn}{\textit}
\newcommand{\Ord}{\mathrm{O}}
\newcommand{\from}{\leftarrow}
\newcommand{\En}{\mathcal{N}}

\begin{document}

\title{A message passing approach for general epidemic models}
\author{Brian Karrer}
\affiliation{Department of Physics, University of Michigan, Ann Arbor, MI
48109}
\author{M. E. J. Newman}
\affiliation{Department of Physics, University of Michigan, Ann Arbor, MI
48109}
\affiliation{Center for the Study of Complex Systems, University of
Michigan, Ann Arbor, MI 48109}

\begin{abstract}
  In most models of the spread of disease over contact networks it is
  assumed that the probabilities per unit time of disease transmission and
  recovery from disease are constant, implying exponential distributions of
  the time intervals for transmission and recovery.  Time intervals for
  real diseases, however, have distributions that in most cases are far
  from exponential, which leads to disagreements, both qualitative and
  quantitative, with the models.  In this paper, we study a generalized
  version of the SIR (susceptible-infected-recovered) model of epidemic
  disease that allows for arbitrary distributions of transmission and
  recovery times.  Standard differential equation approaches cannot be used
  for this generalized model, but we show that the problem can be
  reformulated as a time-dependent message passing calculation on the
  appropriate contact network.  The calculation is exact on trees
  (i.e.,~loopless networks) or locally tree-like networks (such as random
  graphs) in the large system size limit.  On non-tree-like networks we
  show that the calculation gives a rigorous bound on the size of disease
  outbreaks.  We demonstrate the method with applications to two specific
  models and the results compare favorably with numerical simulations.
\end{abstract}

\maketitle

\section{Introduction}
The mathematical modeling of infectious disease outbreaks in human
populations has a long history, stretching back to the pioneering work of
Lowell Reed, Anderson McKendrick, and others in the early twentieth
century~\cite{AM91}.  The standard analytic approach involves dividing the
modeled population into classes or \textit{compartments} according to their
status with respect to the disease of interest---uninfected but
susceptible, infected, recovered, and so forth---and then writing
differential equations to describe the mass flow of individuals between
compartments according to the dynamics of the infection
process~\cite{AM91,Hethcote00}.

Such compartmental models have proven flexible, tractable, and highly
informative as a general guide to the population-level behavior of
diseases, but they also suffer from a number of serious deficiencies, of
which two are particularly significant.  The first, which has attracted a
lot of recent attention in the literature, is the assumption of random
mixing.  In order to write differential equations for flows between
compartments, we must make a \textit{fully mixed} or \textit{mass-action
  approximation} whereby we assume that the probability of disease-causing
contact with any member of a particular compartment is the same.  In real
life this is far from true---most people have high probability of contact
with only that small fraction of the population they rub shoulders with
regularly, and a very small chance of contact with everyone else.  The
incorporation of more realistic mixing patterns into epidemiological
modeling has given rise to the field of \textit{network epidemiology}, in
which contacts are modeled as a network, either static~\cite{PV01a, PV01b,
  MPV02, Keeling19971, Keeling20021, Sharkey20081, Volz20081} or
dynamic~\cite{Volz20071,Gross06,Read08}, and the structure of the network
can have a profound impact on the spread of the disease~\cite{Keeling01,
  Barthelemy20051, Trapman20071, Bansal20071}.

In this paper, however, we focus on a different shortcoming of
compartmental models, one that has by comparison received little attention,
but which is at least as important as the mass-action approximation.  In
order to write down the differential equations of a compartmental disease
model, one must make the assumption that movement between compartments
takes place at a stochastically constant rate.  In modeling a disease from
which most victims recover, for instance, one typically assumes that an
infected individual has a constant probability per unit time of recovery.
While being a useful assumption from a mathematical point of view, however,
this behavior is very far from that of most real diseases.  The assumption
of constant probability of recovery implies an exponential distribution of
times for which individuals remain infected, so that the most probable
duration of infection is zero, and probability decreases uniformly with
time.  In reality, most diseases show a roughly constant duration of
infection---a week, say, or a month---with relatively small fluctuations
from person to person, so that the distribution of durations has a sharp
peak about the average value and is highly nonexponential.  Such
nonexponential distributions are known to have a substantial effect, both
qualitative and quantitative, on the shape of
epidemics~\cite{Lloyd200159,Lloyd01,Wearing05,Vasquez07,Iribarren2009}.

If one is willing to make the mass-action approximation, then
nonexponential behavior can be incorporated into epidemic models by
reformulating the theory in terms of integro-differential
equations~\cite{HT80,KG97}.  If, however, one wishes also to retain the
advances of network epidemiology in representing nonrandom contact
patterns, then even this approach does not work and a new method of
solution is necessary.  In this paper we demonstrate that in the latter
case the calculations can be reformulated in the language of message
passing algorithms of the kind known as belief propagation or sum-product
methods.  In addition to providing exact solutions for the dynamics of
quite general epidemic models on large classes of networks, the message
passing formulation also leads to a number of other results concerning
network epidemiology, including a rigorous upper bound on the size of
disease outbreaks, results for late-time behavior, and results for the
average behavior of epidemics in random network ensembles.

\section{A message passing formulation of epidemics}
We begin by defining the problem.  We consider the simplest nontrivial
model of epidemic disease, the SIR model, in which an individual can be in
one of three disease states, susceptible, infected, or recovered.  We will
assume an initial condition for the epidemic in which each vertex is
susceptible with independent probability~$z$ and infected otherwise.

We assume that disease transmission is taking place on a given contact
network, meaning that disease can only be transmitted between individuals
who are directly connected by an edge in that network.  We also generalize
the model to allow for nonexponential distributions of the times at which
transitions between these states occur, i.e.,~the times at which infection
and recovery occur.  To be completely general, let us define
$s(\tau)\>\d\tau$ to be the probability that an individual infected with
the disease of interest first makes contact sufficient to transmit the
disease to a particular network neighbor at a time between $\tau$ and
$\tau+\d\tau$ after their infection.  Similarly let us define
$r(\tau)\>\d\tau$ to be the probability that an individual infected with
the disease recovers from it at a time between $\tau$ and $\tau+\d\tau$
after infection.

An infected individual can only transmit the disease to a susceptible
neighbor if they are still infected at the time of contact, and hence the
probability that transmission actually occurs between $\tau$ and
$\tau+\d\tau$ after infection is equal to the probability $s(\tau)\>\d\tau$
times the probability $\int_\tau^\infty r(\tau')\>\d\tau'$ that the
individual has not yet recovered.  Let us denote this overall probability
of transmission by $f(\tau)\>\d\tau$:
\begin{equation}
f(\tau)\>\d\tau = s(\tau)\>\d\tau \int_\tau^\infty r(\tau') \>\d\tau'.
\label{eq:foftau}
\end{equation}
Note that this function, unlike $s(\tau)$ and $r(\tau)$ does not integrate
to unity.  Rather, it integrates to the total probability that a vertex
transmits the disease to its neighbor before it recovers, a probability
referred to elsewhere variously as the \textit{transmissibility}
or \textit{infectivity} of the disease.

\begin{figure}
\includegraphics[width=\columnwidth]{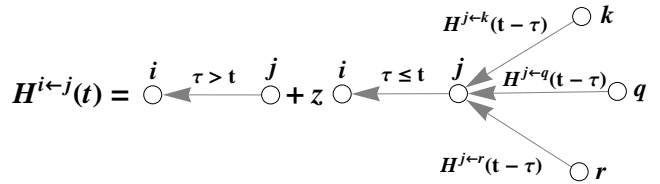}
\caption{The probability that vertex~$i$ does not contract the disease from
  its neighbor~$j$ before time~$t$ is equal to the probability that $j$
  fails to transmit the disease within an interval~$t$ of catching it, plus
  the probability that it does transmit the disease within an interval~$t$
  but that $j$ received the disease from \emph{its} neighbors (here denoted
  $k$, $q$, and~$r$) too late to pass it on to~$i$ in time.}
\label{fig:example}
\end{figure}

The fundamental quantity appearing in our message passing formulation of
disease transmission---the ``message'' that is passed among network
vertices in the calculation---is the probability, which we denote
$H^{i\from j}(t)$, that a vertex~$j$ has \emph{not} passed the disease to
neighboring vertex~$i$ by absolute time~$t$.  (Without loss of generality,
we will assume the epidemic to begin at absolute time~$t=0$.)  An
especially simple case of our approach arises when the network of interest
takes the form of a tree, i.e.,~a network having no loops.  In this case,
the failure of vertex~$j$ to pass the disease to vertex~$i$ before time~$t$
can occur in either of two ways, as illustrated in Fig.~\ref{fig:example}.
First, it may be that, if and when vertex~$j$ contracts the disease, it
fails to transmit it to~$i$ within an interval~$t$ from infection, in which
case clearly $i$ does not contract the disease before absolute time~$t$.
The probability of this occurrence is $1-\int_0^t f(\tau)\>\d\tau$.

The second possibility is that $j$ is scheduled to transmit the disease
within time~$t$ of contracting it, but that $j$ itself got the disease
(from one of its other neighbors) too late for that transmission to occur
before absolute time~$t$, or indeed never got the disease at all.  If $j$
transmits the disease at time~$\tau$ after contracting it, but fails to
contract the disease before time~$t'=t-\tau$ then $i$ does not receive the
disease before time~$t$.  The probability that $j$ does not contract the
disease before~$t'$ is $z\prod_{l \in \En(j)\backslash i} H^{j \from
  l}(t')$, where the leading factor of~$z$ represents the probability that
$j$ was not one of those vertices initially infected with the disease at
$t=0$.  The notation $\En(j)\backslash i$ denotes the set of neighbors of
vertex~$j$, excluding vertex~$i$.  Now integrating over~$t'$, we find the
total probability that $j$ fails to transmit the disease before~$t$ to be
$z \int_0^t f(t-t') \prod_{l \in \En(j)\backslash i} H^{j\from l}(t') \>\d
t'$.

Putting the two contributions to $H^{i\from j}(t)$ together and writing
$t-t'=\tau$ we arrive at the message passing equation
\begin{equation}
H^{i\from j}(t) = 1 - \int_0^t\! f(\tau) \biggl[ 1 -
          z \!\! \prod_{l \in \En(j)\backslash i} \!\!
          H^{j\from l}(t-\tau) \biggr] \>\d\tau.
\label{eq:mptree}
\end{equation}
For the special case of a network taking the form of a tree, this equation
gives us, at least in principle, a complete solution for the probabilities
$H^{i\from j}(t)$ for all~$t$ and arbitrary~$f(\tau)$.

Normally, however, $H^{i\from j}$ is not the quantity of epidemiological
interest.  More commonly one wants to know things such as the fraction of
the population that will be in the various disease states at different
times, or more generally the probability that a particular individual will
be in each disease state.  Let us define $S_i(t)$ to be~1 if individual~$i$
is susceptible at time~$t$ and 0 otherwise, and similarly define $I_i(t)$
and~$R_i(t)$ for the infected and recovered states.  Then $P(S_i(t)=1)$
denotes the probability that vertex~$i$ is susceptible at time~$t$.  For
the sake of economy we will also write this probability more briefly simply
as~$P(S_i)$.  For $i$ to be susceptible at time~$t$ we require (a)~that $i$
is not one of the vertices initially infected at $t=0$, which happens with
probability~$z$, and (b)~that $i$ not receive the infection from any of its
neighbors before time~$t$.  Thus $P(S_i)$ can be expressed quite simply as
\begin{equation}
P(S_i) = z \!\! \prod_{j \in \En(i)} \!\! H^{i\from j}(t).
\label{eq:treepsi}
\end{equation}

Once we have $P(S_i)$, however, one can also immediately calculate $P(I_i)$
and~$P(R_i)$.  Note that the rate $\d P(I_i)/\d t$ at which $P(I_i)$
increases is equal to the rate at which $P(S_i)$ decreases---since all
individuals moving out of the susceptible state must move into the infected
state---minus the rate at which~$i$ recovers.  The recovery rate has two
contributions: the probability~$1-z$ that~$i$ was infected at time~$t=0$
times the rate $r(t)$ of recovery a time~$t$ later, and the probability
that $i$ was infected at some later time~$t'<t$ (which is simply $-\d
P(S_i)/\d t'$) times the rate $r(t-t')$ of recovery a time $t-t'$ later.
This allows us to write a rate equation for $P(I_i)$ thus:
\begin{equation}
{\d P(I_i)\over\d t} = - {\d P(S_i)\over\d t} - (1-z)\,r(t)
                       + \int_0^t r(t-t') {\d P(S_i)\over\d t'} \>\d t'.
\end{equation}
By integrating this equation we can calculate $P(I_i)$ for any~$t$, and
then we can calculate $P(R_i)$ from the knowledge that the probabilities of
the three states must sum to one:
\begin{equation}
P(R_i) = 1 - P(S_i) - P(I_i).
\label{eq:pri}
\end{equation}
Between them, Eqs.~\eqref{eq:mptree} to~\eqref{eq:pri} now give us a
complete solution for the three probabilities, for arbitrary (including
nonexponential) distributions of infection and recovery times.

\section{Message passing on non-tree networks}
The developments of the previous section give us a solution for the SIR
model in the case where the network of interest has no loops, but almost
all real-world networks have loops, and usually many of them.  It is known
that message passing methods, while not exact on non-tree networks can
still give good approximate answers in some cases.  In the present case,
however, we can go further than such qualitative statements and show that
our message passing calculation provides a rigorous upper bound to the
number of infected individuals on networks that contain loops.  To prove
this result, consider the following alternative formulation of the epidemic
process.

In the generalized SIR model discussed here, evolution of the disease
involves infected individuals spreading infection to their susceptible
neighbors at times after infection drawn from the distribution~$s(\tau)$
and recovering at times after infection drawn from~$r(\tau)$.  There is,
however, no requirement that we draw these times at the moment of
infection.  We can if we wish draw them ahead of time, before executing the
steps of the model.  That is, we can for each vertex~$i$ in the network
draw a time~$\tau_i$ from the distribution~$r(\tau)$ and associate it with
that vertex.  When vertex~$i$ becomes infected, we look up the value
of~$\tau_i$ which tells us the interval of time before~$i$ recovers.  For
the edges the situation is only a little more complicated.  We replace each
undirected edge in the network with two directed edges pointing in opposite
directions, to represent the act of disease transmission in either
direction between the two relevant vertices.  Then for each directed edge
$j \to i$ we draw a time~$w_{ij}$ from the distribution~$s(w)$ to represent
the time after infection of~$j$ at which contact is made with~$i$.  If this
time falls before the recovery of~$j$, i.e.,~if $w_{ij}<\tau_j$, then
transmission will take place if $j$ is ever infected, and will occur an
interval~$w_{ij}$ after infection.  If, however, $j$~recovers first,
i.e.,~if $w_{ij}>\tau_j$, then no transmission takes place, which we can,
if we wish, represent mathematically by setting~$w_{ij}=\infty$.

The end result is a directed ``transmission network'' in which the edges
represent possible transmission events and the values~$w_{ij}$ on the edges
represent the time delay between arrival of the infection at~$j$ (if that
ever happens) and arrival of the infection at~$i$.

In terms of this network it is now quite simple to write down the
probability $P(S_i)$ that vertex~$i$ is susceptible at time~$t$.  In order
to be susceptible we require (a)~that $i$ was not infected at time~$0$,
which happens with probability~$z$, and (b)~that there exists no path from
any vertex~$j$ to vertex~$i$ such that vertex~$j$ was infected at
time~$0$ and the sum of the time delays~$w_{ij}$ along the path is less
than~$t$.

An alternative way of thinking about this second condition is to consider
the neighborhood of radius~$t$ about vertex~$i$, meaning the set of
vertices~$j$ a distance~$t$ or less from~$i$, where distance is measured in
terms of the sum of the values~$w_{ij}$ along the path---the shortest
weighted distance in the language of graph theory.  If any of the vertices
in this neighborhood is infected at time zero then vertex~$i$ will not be
susceptible at time~$t$.  Let us suppose that there are $n_i$ vertices in
the neighborhood, excluding vertex~$i$ itself.  Then the probability
that~$i$ is susceptible---for this particular choice of the~$w_{ij}$
and~$\tau_i$---is $z^{n_i+1}$.  We are interested, however, in the
probability averaged over all values of the~$w_{ij}$ and~$\tau_i$, which is
\begin{equation}
P(S_i) = z \av{z^{n_i}},
\end{equation}
where the angle brackets $\av{\ldots}$ denote the average over the ensemble
of values of~$w_{ij}$ and~$\tau_i$.

This equation is correct and exact in all cases.  To relate it to our
previous message passing approach and understand how the calculation
proceeds on networks with loops, consider the alternative set of vertex
counts~$n_{ij}$, which are the numbers of vertices whose distance to~$i$ is
$t$ or less, but now with the restriction that the penultimate vertex along
the path to~$i$ must be vertex~$j$.  For reasons that will
shortly become clear, we also forbid paths that pass through vertex~$i$
more than once.  That is, there may be a path of length~$t$ or less that
first passes through~$i$ to reach~$j$ and then returns to~$i$, but such
paths are disallowed.  In practice, a simple way to enforce this constraint
is to remove from the network all directed edges outgoing from vertex~$i$.
In this case, vertex~$i$ is said to be a \defn{cavity vertex} or \defn{in
the cavity state}.

\begin{figure}
\includegraphics[width=9cm]{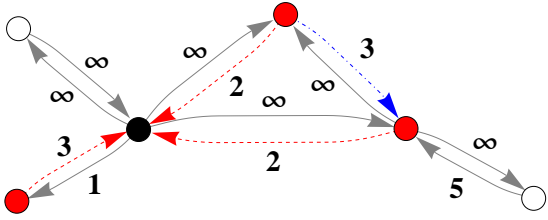}
\caption{A small directed transmission network in which each edge is
  labeled with its associated transmission delay~$w_{ij}$, except for edges
  with $w_{ij} > \tau_j$, which are labeled~$\infty$.  The three red
  vertices denote those within distance~$6$ of the black vertex and the red
  edges correspond to the weighted shortest paths from the red vertices to
  the black one.  If we approximate the number of red vertices as in
  Eq.~\eqref{eq:nontreeapprox} by the sum of the numbers of vertices within
  distance~$6$ that are reachable via each of the black vertex's immediate
  neighbors, then we will count four vertices instead of three: the top red
  vertex will be counted twice because the blue edge provides a second path
  from this vertex to the black one.}
\label{fig:graphfig}
\end{figure}

We now observe that, as illustrated in Fig.~\ref{fig:graphfig}, the sum of
$n_{ij}$ over all neighbors~$j$ is always at least as great as~$n_i$:
\begin{equation}
n_i \le \sum_{j \in \En(i)} n_{ij},
\label{eq:nontreeapprox}
\end{equation}
where the inequality becomes an exact equality if the network is a tree.
(It is in order to ensure this equality that we exclude paths that pass
through~$i$ twice.)  Then $z^{n_i} \ge z^{\sum_{j \in \En(i)} n_{ij}}$
and
\begin{equation}
P(S_i) = z \av{z^{n_i}} \ge z \bigl\langle z^{\sum_{j \in \En(i)}
         n_{ij}} \bigr\rangle
       = z \Bigl\langle \prod_{j \in \En(i)} z^{n_{ij}} \Bigr\rangle.
\label{eq:ineq1}
\end{equation}

We now apply a version of the Chebyshev integral inequality, proved in the
appendix, that for any set of non-negative functions $f_1(x_1,\ldots,x_k),
\ldots, f_n(x_1,\ldots,x_k)$ that are monotone increasing or decreasing in
every argument, says
\begin{align}
\biggl\langle \prod_{i=1}^n f_i(x_1,\ldots,x_k) \biggr\rangle
  \ge \prod_{i=1}^n \bigl\langle f_i(x_1,\ldots,x_k) \bigr\rangle,
\end{align}
where the average is over any distribution of independent
variables~$x_1,\ldots,x_k$.  Applied to Eq.~\eqref{eq:ineq1}, this
inequality tells us that
\begin{equation}
P(S_i) \ge z \!\! \prod_{j \in \En(i)} \! \bigl\langle z^{n_{ij}} \bigr\rangle
       = z \!\! \prod_{j \in \En(i)} \!\! H^{i\from j}(t),
\label{eq:nontreepsi}
\end{equation}
where we have defined
\begin{equation}
H^{i\from j}(t) = \bigl\langle z^{n_{ij}} \bigr\rangle.
\end{equation}
This quantity is the average probability that at time~$t$ the infection has
not been passed to vertex~$i$ via neighbor~$j$ (again excluding cases where
the infection passes through~$i$ twice).  It plays the same role as the
corresponding quantity in Eq.~\eqref{eq:treepsi} for the case of a tree,
and we can evaluate it in an analogous way.  As before we split $H^{i\from
  j}(t)$ into two parts.  The first is the probability that, even if $j$ is
infected, it does not transmit the disease to~$i$ within time~$t$ of
infection.  This probability, as before, is $1 - \int_0^t f(\tau)\>\d\tau$,
where $f(\tau)$ is defined by Eq.~\eqref{eq:foftau}.

The second part is the probability that $j$ is scheduled to transmit the
disease within time~$\tau<t$ of contracting it, but that $j$ itself gets
the disease too late for the transmission to occur before absolute time~$t$
(or $j$ never gets the disease at all).  For transmission before time~$t$
vertex~$j$ needs to contract the disease before $t'=t-\tau$ and the
probability that this does not happen is $P(S_j(t')|\mbox{$i$ in
  cavity})$, where it is now important that~$i$ is in the cavity state, so
as to disallow paths for infection that pass through~$i$ itself.  Then the
probability that $j$ fails to transmit the disease before time~$t$ is
$\int_0^t f(t-t') P(S_j(t')|\mbox{$i$ in cavity})\>\d t'$.

The probability $P(S_j(t')|\mbox{$i$ in cavity})$ we can calculate from the
appropriate analog of Eq.~\eqref{eq:nontreepsi} but with both~$i$ and $j$
in the cavity state, i.e.,~with their outgoing edges deleted.  But consider
now adding back in all the edges leading from~$i$ except the one to~$j$.
In doing so we only add paths to the network and hence potentially increase
the size of the neighborhood of vertex~$j$ but never decrease it.  This
implies that we only decrease~$P(S_j(t'))$, so that
\begin{align}
P(S_j(t')|\mbox{$i$ in cavity}) &\ge P(S_j(t')|\mbox{$i\to j$ deleted})
       \nonumber\\
  &\ge z \!\! \prod_{l \in \En(j)\backslash i} \!\!
              \bigl\langle z^{n_{jl}} \bigr\rangle
       \nonumber\\
  &=   z \!\! \prod_{l \in \En(j)\backslash i} \!\!\! H^{j\from l}(t'),
\end{align}
where we have used Eq.~\eqref{eq:nontreepsi}.  Combining our two
contributions to $H^{i\from j}(t)$ and writing $t-t'=\tau$, we now find
that
\begin{equation}
H^{i\from j}(t) \ge 1 - \int_0^t\! f(\tau) \biggl[
  1 - z \!\! \prod_{l \in \En(j)\backslash i} \!\!\!
      H^{j\from l}(t-\tau) \biggr] \>\d\tau.
\label{eq:finalineq}
\end{equation}

This result is very similar to the message passing equality of
Eq.~\eqref{eq:mptree}, but it is an inequality, and hence cannot be
directly employed to calculate properties of the epidemic.  Let us,
however, define a different function~$F^{i\from j}(t)$ by the equation
\begin{equation}
F^{i\from j}(t) = 1 - \int_0^t\! f(\tau) \biggl[
  1 - z \!\! \prod_{l \in \En(j)\backslash i} \!\!\!
      F^{j\from l}(t-\tau) \biggr] \>\d\tau,
\label{eq:finalmp}
\end{equation}
which is an equality and so \emph{can} be used to calculate~$F^{i\from j}$,
for instance by iteration starting from a suitable initial value~$F^{i\from
  j}_0(t)$.  Suppose we choose as our initial value $F^{i\from j}_0(t) =
H^{i\from j}(t)$ for all $i,j$ and~$t$, so that, from
Eq.~\eqref{eq:finalineq}, we have
\begin{equation}
F^{i\from j}_0(t) \ge 1 - \int_0^t\! f(\tau) \biggl[
  1 - z \!\! \prod_{l \in \En(j)\backslash i} \!\!\!
      F^{j\from l}_0(t-\tau) \biggr] \>\d\tau.
\label{eq:g0ineq}
\end{equation}
(Of course we don't know the value of $H^{i\from j}(t)$, but suppose for
the moment that we do.)  Then, performing one step of the iteration, we
arrive at a new value~$F^{i\from j}_1(t)$ thus:
\begin{align}
F^{i\from j}_1(t) &= 1 - \int_0^t\! f(\tau) \biggl[
  1 - z \!\! \prod_{l \in \En(j)\backslash i} \!\!\!
      F^{j\from l}_0(t-\tau) \biggr] \>\d\tau
  \nonumber\\
  &\le F^{i\from j}_0(t),
\label{eq:g1def}
\end{align}
where we have used Eq.~\eqref{eq:g0ineq}.  But note that, since
$f(\tau)\ge0$ for all~$\tau$, Eq.~\eqref{eq:g1def} also implies that
\begin{align}
& \int_0^t\! f(\tau) \!\! \prod_{l\in \En(j)\backslash i} \!\!\!
                        F^{j\from l}_1(t-\tau) \>\d\tau \nonumber\\
  &\hspace{5em} \le \int_0^t\! f(\tau) \!\! \prod_{l\in \En(j)\backslash i}
                \!\!\! F^{j\from l}_0(t-\tau) \>\d\tau,
\end{align}
and hence from Eq.~\eqref{eq:g1def}
\begin{equation}
F^{i\from j}_1(t) \ge 1 - \int_0^t\! f(\tau) \biggl[
  1 - z \!\! \prod_{l\in \En(j)\backslash i} \!\!\!
      F^{j\from l}_1(t-\tau) \biggr] \>\d\tau,
\end{equation}
which is the equivalent of Eq.~\eqref{eq:g0ineq} for~$F^{i\from j}_1(t)$.
Now we can repeat the same argument to show that for a general step of the
iteration we must have
\begin{equation}
F^{i\from j}_m(t) \le F^{i\from j}_{m-1}(t).
\end{equation}
In the limit $m\to\infty$, the iteration must converge, since $F^{j\from
  l}_m(t)$ is bounded below by $1 - \int_0^t f(\tau)\>\d\tau$, and hence in
this limit we get a solution to Eq.~\eqref{eq:finalmp} that satisfies
\begin{equation}
F^{i\from j}(t) \le F^{i\from j}_0(t) = H^{i\from j}(t),
\end{equation}
for all $i,j$ and~$t$.

Now, making use of Eq.~\eqref{eq:nontreepsi}, we have
\begin{equation}
P(S_i) \ge z \!\! \prod_{j \in \En(i)} \!\! H^{i\from j}(t)
  \ge z \!\! \prod_{j \in \En(i)} \!\! F^{i\from j}(t).
\label{eq:finalpsi}
\end{equation}
Thus Eq.~\eqref{eq:finalmp} allows us to calculate a rigorous lower bound
on the probability that any vertex is in the susceptible state.  Notice
that Eq.~\eqref{eq:finalmp} is the same as the equation for~$H^{i\from j}$
in the tree case, Eq.~\eqref{eq:mptree}, but is perfectly well defined for
any network, tree or otherwise.

Our lower bound on $P(S_i)$ also gives us upper bounds on $P(I_i)$ and
$P(R_i)$, both of which are trivially less than $1-P(S_i)$, as well as an
upper bound on the sum $P(I_i)+P(R_i)=1-P(S_i)$, which is the total
probability that $i$ has ever caught the disease.  Hence our
message passing calculation can in this case give us an upper bound on the
number of individuals infected by an epidemic, a result of possible
value---a guarantee that infection will not rise above a certain level
could be used as a quality function to quantify the efficacy of proposed
vaccination campaigns or other public health interventions.

Employing Eqs.~\eqref{eq:finalmp} and~\eqref{eq:finalpsi} in a message
passing algorithm would involve propagating messages that take the form of
functions $F^{i\from j}(t)$ of time.  On a tree, one would start with the
leaves of the tree, for which Eq.~\eqref{eq:finalmp} is trivial, and work
inwards through the network until the functions on all edges have been
evaluated.  On a non-tree network, the calculation is more complicated
because one does not in general know any of the $F^{i\from j}(t)$ to begin
with, so one would have to make an initial guess and then iterate
Eq.~\eqref{eq:finalmp} repeatedly to reach convergence.  $F^{i\from
  j}(t)=1$ for all $i,j$ and~$t$ is a suitable starting condition, but the
iteration itself can in practice be time-consuming and the calculation may
not be tractable.  Even if it is tractable, it almost certainly demands
more effort than simply simulating the spread of an epidemic on the network
of interest.  There are some choices of the distributions $r(\tau),s(\tau)$
for which the equations simplify and are more tractable---we examine two in
Section~\ref{sec:cm}.  Alternatively, one may be able to make useful
approximations in some cases.  For instance, if $f(\tau)$~is sharply peaked
close to $\tau=0$, as it is for many real diseases, then it may be
reasonable to approximate $F^{i\from j}(t-\tau)$ in Eq.~\eqref{eq:finalmp}
by its value~$F^{i\from j}(t)$ at $\tau=0$.  Then~\eqref{eq:finalmp}
becomes
\begin{equation}
F^{i\from j}(t) = 1 - p(t) + zp(t) \!\!\! \prod_{l \in \En(j)\backslash i}
                  \!\!\! F^{j\from l}(t),
\end{equation}
where $p(t) = \int_0^t f(\tau)\>\d\tau$.  Hence the values of $F^{i\from
  j}$ at different times decouple and the equations can be solved by a
simple scalar iteration---no integrals need be performed.  Although
efficient, however, this approximation is usually only a good one in
regions where $F^{i\from j}(t)$ is relatively constant over the timescales
typical of the disease progression represented by~$f(\tau)$, which means
early and late times, but not in the crucial intermediate interval where
most of the interesting behavior falls.

Even in cases where the message passing equations are not a practical
calculational tool, however, they can still be useful.  In particular, they
can tell us about the late-time limit of epidemics, including important
quantities such as the total number of people infected by the disease, and
they allow us to calculate epidemic outcomes averaged over ensembles of
networks such as the widely-studied configuration model.  We look at these
two applications now in turn.

\section{Late-time behavior}
Taking the limit $t\to\infty$ in Eq.~\eqref{eq:finalmp}, we get
\begin{equation}
F^{i\from j}(\infty) = 1 - \int_0^\infty\! f(\tau) \biggl[
  1 - z \!\! \prod_{l\in \En(j)\backslash i} \!\!\!
      F^{j\from l}(\infty) \biggr] \>\d\tau,
\end{equation}
where we have assumed that $f(\tau)$ is suitably small for large values of
its argument.  Writing $F^{i\from j}=F^{i\from j}(\infty)$ for short
and defining $p=\int_0^\infty f(\tau)\>\d\tau$, which is the total
probability of transmission occurring between two vertices connected by an
edge, we then find that
\begin{equation}
F^{i\from j} = 1 - p + p z \!\! \prod_{l\in \En(j)\backslash i} \!\!\!
                       F^{j\from l}.
\label{eq:latetime1}
\end{equation}
This again takes the form of a message passing calculation, but now the
messages passed are simple numbers, and hence the calculation can be
performed quickly, even on networks that are not trees.  Then the
probability that a vertex is susceptible in the limit of late times
satisfies
\begin{equation}
P(S_i) \ge z \!\! \prod_{j \in \En(i)} \!\! F^{i\from j}.
\label{eq:latetime2}
\end{equation}
In the limit of late times there are no infected individuals---all have
either recovered or never got sick in the first place---so
$P(R_i)=1-P(S_i)$.  Thus this calculation gives us an upper bound on the
probability that any given individual ever contracts the disease or, if we
sum over all vertices, an upper bound on the size of the disease outbreak.

As has been discussed
previously~\cite{Newman02c,Miller07,Kenah2007,Miller08}, the late-time
limit of the SIR model is related to a correlated bond percolation process
on the corresponding directed transmission network, the correlations
arising because of variation in the time an individual takes to recover: if
an individual recovers quickly then the probability of transmission of the
disease to any of its neighbors is small; if it takes a long time to
recover the probability is correspondingly larger.
Equations~\eqref{eq:latetime1} and~\eqref{eq:latetime2} can be considered
to define a message passing algorithm for solving precisely this bond
percolation problem on a general network.  In this context, $F^{i\from j}$
is a generating function in~$z$ for the number of vertices in the
percolation cluster of vertex~$i$ that are reachable via vertex~$j$, and
$P(S_i)$ is a generating function for the overall sizes of the clusters.
In recent unpublished work, Shiraki and Kabashima~\cite{Shiraki2010} have
given a message passing method for calculating percolation cluster sizes on
trees and locally-tree-like networks, which is equivalent to the method
reported here for the special case of a tree.

\section{Epidemics on configuration model networks}
\label{sec:cm}
Our method can also be used to calculate average probabilities of infection
for ensemble models of networks.  It is common in the study of processes on
networks to look at not the behavior on a single network, but the average
behavior in an ensemble model defined as a probability distribution over
possible networks.  The message passing formalism developed here allows one
to calculate such average behaviors easily.  We demonstrate this type of
calculation using the configuration model, which is probably the most
widely studied ensemble model of a network~\cite{MR95,NSW01}.

In the configuration model one fixes the degree distribution of the
network---meaning the fractions~$p_k$ of vertices with each possible
degree~$k$---but in other respects connects vertices at random.  Thus in
calculating the behavior of an epidemic on the configuration model there
are two sources of randomness to average over.  The first is the randomness
in the dynamics of the disease, which is already built into our
message passing formalism.  The second is the randomness of the graph
ensemble.

Consider the average over the graph ensemble and consider an edge attached
to vertex~$i$.  In different networks of the ensemble this edge will be
attached to different vertices~$j$ at its other end and hence a different
message $H^{i \from j}$ will be transmitted down the edge.  The ensemble
average probability that vertex~$i$ has not yet been infected along the
edge by time $t$ is equal to the average of these messages over the set of
networks.  But, since every edge plays an identical role in the
configuration model ensemble, the average message is the same for all
edges~$i,j$ and hence we need calculate only one message to solve for the
average behavior of the model.  Let us denote this average message
by~$H_1(t)$.

To calculate the average message, we need to average Eq.~\eqref{eq:mptree}
(or its equivalent, Eq.~\eqref{eq:finalmp} for non-tree networks), which
requires us to average the product on the right-hand side of the equation.
The average of such a product is not in general equal to the product of
the average message, which potentially makes the calculation more
complicated.  However, in the limit of large network size, configuration
model networks have the crucial property of being locally tree-like, with
the shortest cycles in the network being of length $\Ord(\log n)$ and hence
diverging as $n\to\infty$.  This means that the messages a vertex receives
along each of its incident edges are independent in the large-$n$
limit---in essence, we assume that correlations along a cycle of diverging
length are irrelevant in the large size limit.  In this case, the average
of the product of messages received by a vertex \emph{is} equal to the
product of the average.

Averaging Eq.~\eqref{eq:mptree} over the ensemble and allowing for the fact
that all messages are the same, the product $\prod_{l \in \En(j)\backslash
  i} H^{j\from l}$ in the equation now becomes simply a power $[H_1(t)]^k$,
where $k$ is the so-called excess degree of~$j$, i.e.,~its degree minus the
edge between $i$ and~$j$, which has been removed because~$i$ is in the
cavity state.  The excess degree is distributed according to the excess
degree distribution $q_k = (k+1)p_{k+1}/\av{k}$~\cite{NSW01} and, averaging
over this distribution, we find
\begin{align}
H_1(t) &= \sum_{k=0}^\infty q_k \biggl[ 1 - \int_0^t\! f(\tau) \Bigl( 1 -
          z \bigl[ H_1(t-\tau) \bigr]^k \Bigr) \>\d\tau \biggr] \nonumber\\
       &= 1 - \int_0^t\! f(\tau) \bigl[ 1 - z G_1(H_1(t-\tau)) \bigr]
          \>\d\tau,
\label{eq:cmh1}
\end{align}
where $G_1(x) = \sum_k q_k x^k$ is the probability generating function for
the excess degree distribution.

Similarly, from Eq.~\eqref{eq:treepsi}, the probability that a vertex of
(ordinary) degree~$k$ is susceptible at time~$t$ is $z [H_1(t)]^k$ and the
average probability of being susceptible is
\begin{equation}
P(S) = z \sum_{k=0}^\infty p_k [H_1(t)]^k = z G_0(H_1(t)),
\label{eq:cmh0}
\end{equation}
where $G_0(x) = \sum_k p_k x^k$ is the generating function for the ordinary
degree distribution~$p_k$.

Again we can study the late-time behavior by letting $t\to\infty$ and
writing $H_1=H_1(\infty)$ to give
\begin{equation}
H_1 = 1 - p + p z G_1(H_1),
\end{equation}
and
\begin{equation}
P(S) = z G_0(H_1),
\end{equation}
where $p = \int_0^\infty f(\tau) \>\d\tau$.  These two equations are
precisely the standard equations for bond percolation on the configuration
model~\cite{CNSW00} and highlight again the connection between the SIR
model and percolation theory.  The message~$H_1$ can be regarded as a
generating function in~$z$ for the distribution of numbers of vertices
reachable along an edge in bond percolation and $P(S)$ is a generating
function for the sizes of clusters.

\begin{figure*}
\begin{center}
\includegraphics[width=15cm]{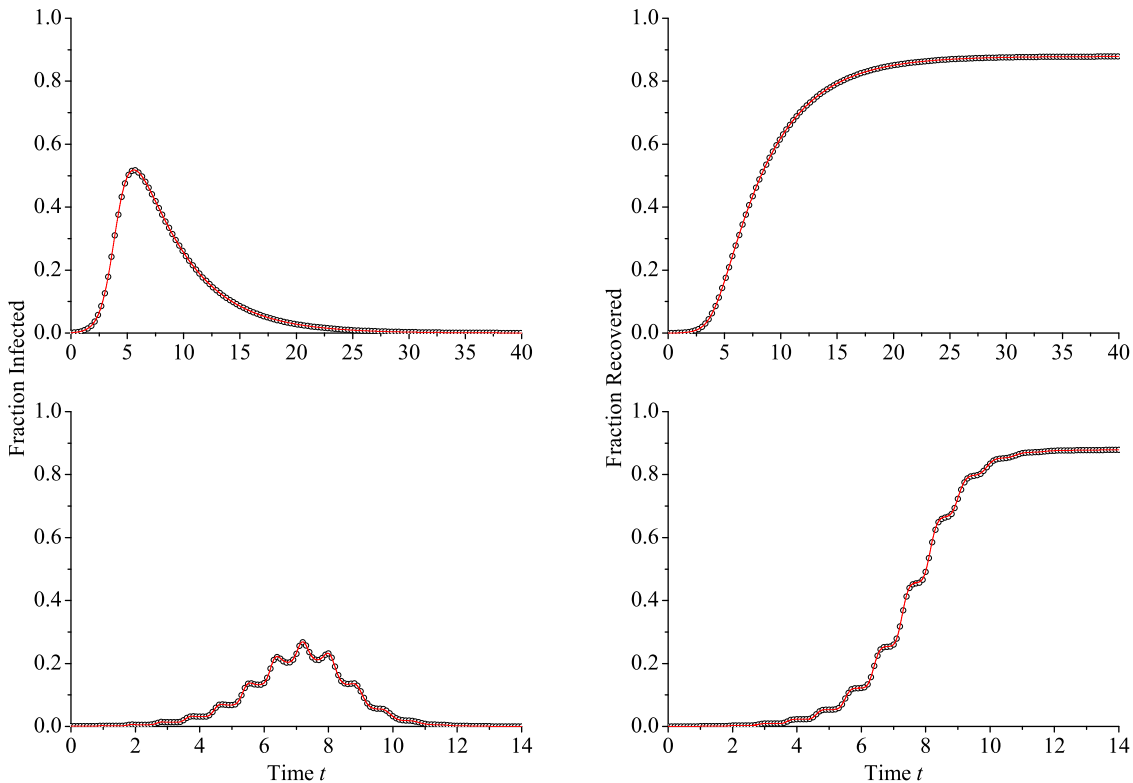}
\end{center}
\caption{Fraction of the population infected (left) and
  recovered (right) as a function of time for two different choices of the
  parameters of the model.  Calculations were performed on configuration
  model networks of $n=10^5$ vertices and Poisson degree distribution with
  mean~3.  In the top two panels infection and recovery times are
  exponentially distributed as described in the text, with $\beta =
  \frac89$ and $\gamma = \frac29$.  In the bottom two panels $f(\tau)$
  takes the ``top hat'' form of Eq.~\eqref{eq:tophat}, with $\tau_s=0.8$,
  $\tau_r =1$, and~$p=0.8$.  The initial condition was $z=0.999$ in each
  case.  Solid lines in each panel are the predictions of the theory;
  circles are simulation results, averaged over $100$ runs.}
\label{fig:poissonfig}
\end{figure*}

\section{Examples}
As a first example of the application of our formalism, consider what
happens when the distributions $r(\tau)$ and $s(\tau)$ take the standard
exponential form, corresponding to stochastically constant probabilities of
infection with and recovery from disease.  Specifically, we assume that
$s(\tau)=\beta \e^{-\beta\tau}$ and $r(\tau) = \gamma \e^{-\gamma\tau}$,
where $\beta$ and $\gamma$ are the rates of infection and recovery.  Then
$f(\tau) = \beta \e^{-(\beta+\gamma)\tau}$ and, making the substitution
$t'=t-\tau$, Eq.~\eqref{eq:cmh1} becomes
\begin{equation}
H_1(t) = 1 - \beta\e^{-(\beta+\gamma)t} \int_0^t \e^{(\beta+\gamma)t'}
         \bigl[ 1 - zG_1(H_1(t')) \bigr] \>\d t'.
\end{equation}
Differentiating with respect to~$t$, we then find that $H_1$ satisfies
\begin{align}
{\d H_1\over\d t} &= \beta(\beta+\gamma)\e^{-(\beta+\gamma)t}
                     \int_0^t \e^{(\beta+\gamma)t'}
                     \bigl[ 1 - zG_1(H_1(t')) \bigr] \>\d t' \nonumber\\
                  &\quad {} - \beta \bigl[ 1 - zG_1(H_1(t)) \bigr]
                  \nonumber\\
                  &= \gamma - (\beta+\gamma) H_1(t) + \beta zG_1(H_1(t)).
\end{align}
with the initial condition $H_1(0)=1$.  This differential equation has the
solution
\begin{equation}
t = \int_1^{H_1} {\d u\over\gamma-(\beta+\gamma)u + \beta zG_1(u)}.
\end{equation}
And once we have $H_1(t)$ we can use Eq.~\eqref{eq:cmh0} to calculate
$P(S)$ and subsequently $P(I)$ and~$P(R)$.  In Fig.~\ref{fig:poissonfig}
(top two frames) we show the form of the resulting solution for the
particular choice of a Poisson degree distribution, along with the results
of numerical simulations of epidemics spreading on the same networks.  As
the figure shows, the analytic and numerical approaches agree well, and
take the familiar form of an SIR outbreak with a brief peak in the number
of infected individuals followed by a sharp decline and corresponding rise
in the number of recovered individuals.

But now consider a second choice that is quite different but perhaps more
realistic.  In this case we assume that individuals once infected do not
become infectious immediately, passing through a latent period before
developing a transmissible infection, and also that infected individuals do
not start recovering from disease immediately upon infection as in the
exponential model, but remain infected for a certain length of time then
recover.  A simple choice displaying these two behaviors is the ``top hat''
function
\begin{equation}
f(\tau) = {p\over\tau_r-\tau_s} \bigl[ \theta(\tau-\tau_s)
                                       - \theta(\tau-\tau_r) \bigr],
\label{eq:tophat}
\end{equation}
with $\tau_r>\tau_s$, where $\theta(\tau)$ is the Heaviside step function.
In this expression $\tau_s$ is the time at which the infected individual
becomes infectious, $\tau_r$~is the time at which they recover, and~$p$, as
before, is the total probability of transmission.

Inserting this form into Eq.~\eqref{eq:cmh1} and again differentiating gives
\begin{align}
{d H_1\over\d t} &= {p\over\tau_r-\tau_s} \bigl[
                    \theta(t-\tau_r) [ 1 - zG_1(H_1(t-\tau_r)) ] \nonumber\\
  &\qquad {} - \theta(t-\tau_s) [ 1 - zG_1(H_1(t-\tau_s)) ] \bigr].
\end{align}   
where again $H_1(0)=1$.  The lower two panels of Fig.~\ref{fig:poissonfig}
show the solution of this equation for the same Poisson degree distribution
as previously, and $p$, $\tau_r$, and~$\tau_s$ chosen so as to give the
same mean time of transmission and total transmission probability as in the
exponential case.  Fixing the total transmission probability to be the same
also fixes the long-time behavior to be the same, as can be seen in the
figure.

The two calculations---exponential and ``top hat'' versions
of~$f(\tau)$---nonetheless give quite different results.  The epidemic
peaks around the same time in each (about $t=6$ in the plots), but more
individuals are infected at any time in the exponential case and the
epidemic lasts longer.  Furthermore, the top hat case shows distinctive
waves of infection, of period roughly equal to~$\tau_s$, separated by
intervals of comparatively lower disease activity.  These waves are caused
by the appearance of distinct ``generations'' in the spread of the disease
as the first round of disease carriers passes infection to the second, who
some time later pass it to the third, and so on.  Such waves of infection
are observed in many real-world diseases but are absent from models using a
conventional exponential distribution of infection times (although they can
be represented in a crude fashion by introducing additional disease states,
as in the so-called SEIR model).

For other choices of degree distribution, including power-law, uniform, and
exponential distributions, the predictions are qualitatively similar by and
large, and agree similarly well with simulation results.  The shapes of the
curves are, however, significantly altered by different choices of the
parameters $\tau_r$ and~$\tau_s$ in the top hat case: as the values of
$\tau_r$ and $\tau_s$ become better separated the waves of infection become
blurred and ultimately impossible to distinguish.  Conversely, the waves
become more pronounced if $\tau_r$ and $\tau_s$ are chosen closer to one
another.

\section{Conclusions}
In this paper, we have studied the SIR model of epidemic disease on a
contact network, in a generalized form that allows for non-constant
probabilities of infection and recovery, by contrast with conventional SIR
calculations.  Abandoning constant probabilities obliges us also to abandon
the traditional differential equation approach to solving the model, but we
have shown that the problem can be reformulated instead in the language of
message passing.  We have given a message passing calculation that is exact
on networks that take the form of trees (or are locally tree-like, as in
random graphs) and provides a rigorous bound on the probabilities of
disease states on non-tree-like networks.

We have demonstrated the application of our approach to the calculation of
the late-time behavior of the generalized SIR model and to the calculation
of average properties of the model within the random graph ensemble known
as the configuration model.  One could in principle extend the calculations
to other random graph ensembles, such as random graphs with degree
correlations~\cite{Newman02f} or random graphs with
clustering~\cite{Newman09b}, or to calculations on single networks
(i.e.,~not ensembles).

The approach taken here can be applied to other dynamical models on
networks, such as the SI or SEIR models, again yielding exact results on
trees or tree-like networks and rigorous bounds in the non-tree case, and
it is possible the approach could also be applied to threshold
models~\cite{Dodds04}.  At the moment, it's unclear whether models such as
the SIS model in which vertices can return to past states can be tackled in
the message passing framework.  The developments for the SIR model relied
on our having an exact message passing solution on a tree.  We have not yet
been able to find a similar solution for the SIS model and so the
development of a message passing method for this model remains an open
problem.

\begin{acknowledgments}
  The authors thank Lenka Zdeborova for useful conversations.  This work
  was funded in part by the National Science Foundation under grant
  DMS--0804778 and by the James S. McDonnell Foundation.
\end{acknowledgments}

\appendix*
\section{Chebyshev Integral Inequality}
Let $f_1(x_1,\ldots,x_k),\ldots, f_n(x_1,\ldots,x_k)$ be a set of $n$
non-negative functions that are monotone decreasing or increasing in each
of their $k$ real-valued arguments for fixed values of the other arguments.
(They can be increasing in one argument and decreasing in another.)  Then
it can be proved that
\begin{equation}
\biggl\langle \prod_{i=1}^n f_i(x_1,\ldots,x_k) \biggr\rangle
  \ge \prod_{i=1}^n \bigl\langle f_i(x_1,\ldots,x_k) \bigr\rangle,
\end{equation}
where the average is over any distribution of the independent
variables~$x_1,\ldots,x_k$.  The proof is as follows.

Let $\bigl\langle f \bigr\rangle_{x_1\ldots x_j}$ denote the partial
average
\begin{equation}
\int f(x_1,\ldots,x_j,x_{j+1},\ldots,x_k)P(x_1)\ldots P(x_j)\>\d x_1
  \ldots \d x_j,
\end{equation}
which is a function of the remaining arguments $x_{j+1}$ to~$x_k$.  Then
consider the following product for arbitrary $x$ and $y$
\begin{align}
& \bigl[ f_1(x, x_2,\ldots, x_n) - f_1(y, x_2, \ldots, x_n) \bigr]
  \nonumber\\
& \times \biggl[\,\prod_{i=2}^n f_i(x, x_2,\ldots, x_n) - \prod_{i=2}^n
  f_i(y, x_2, \ldots, x_n) \biggr].
\label{eq:chebyshev1}
\end{align}
Because the functions~$f_i$ are non-negative and monotonic in their first
argument, the factors in brackets $[\ldots]$ are either both positive or
both negative and hence the entire expression is non-negative for any $x$
and~$y$.  Now let $x$ and $y$ be independent random variables, both with
the same distribution as~$x_1$ and let us average~\eqref{eq:chebyshev1}
over $x$ and~$y$.  After rearranging we find that
\begin{equation}
\biggr\langle \prod_{i=1}^n f_i \biggr\rangle_{x_1} \!
  \ge \, \bigl\langle f_1 \bigr\rangle_{x_1} \biggr\langle
      \prod_{i=2}^n f_i \biggr\rangle_{x_1}.
\end{equation}
The same argument can now be applied to the remaining functions $f_2,
\ldots, f_n$ in turn, to demonstrate that
\begin{equation}
\biggr\langle \prod_{i=1}^n f_i \biggr\rangle_{x_1} \!
  \ge \,\prod_{i=1}^n \bigl\langle f_i \bigr\rangle_{x_1},
\label{eq:chebyshev2}
\end{equation}
and the equivalent result naturally holds for averages over any of the
variables:
\begin{equation}
\biggr\langle \prod_{i=1}^n f_i \biggr\rangle_{x_j} \!
  \ge \,\prod_{i=1}^n \bigl\langle f_i \bigr\rangle_{x_j},
\label{eq:chebyshev3}
\end{equation}

The remainder of the proof proceeds by induction.  Assume that
\begin{equation}
\biggr\langle \prod_{i=1}^n f_i \biggr\rangle_{x_1\ldots x_j}
  \ge \,\prod_{i=1}^n \bigl\langle f_i \bigr\rangle_{x_1\ldots x_j}
\end{equation}
for $j<k$.  Averaging both sides over one additional variable~$x_{j+1}$
gives
\begin{equation}
\biggr\langle \prod_{i=1}^n f_i \biggr\rangle_{x_1\ldots x_j, x_{j+1}} \!
  \ge \,\,\biggr\langle \prod_{i=1}^n
  \bigl\langle f_i \bigr\rangle_{x_1\ldots x_j} \biggr\rangle_{x_{j+1}}.
\end{equation}
But $\bigl\langle f_1 \bigr\rangle_{x_1\ldots x_j},\ldots,\bigl\langle f_n
\bigr\rangle_{x_1\ldots x_j}$ is itself a set of monotone non-negative
functions of the variables $x_{j+1},\ldots, x_k$.  Applying
Eq.~\eqref{eq:chebyshev3} to this set, we then find that
\begin{equation}
\biggr\langle \prod_{i=1}^n f_i \biggr\rangle_{x_1\ldots x_{j+1}} \!
  \ge \,\prod_{i=1}^n \bigl\langle f_i
  \bigr\rangle_{x_1\ldots x_{j+1}}.
\end{equation}
Applying induction and using Eq.~\eqref{eq:chebyshev2} as the base case,
the result is now established.

\end{document}